\documentclass[preprint]{vgtc}               

\ifpdf
  \pdfoutput=1\relax                   
  \usepackage{graphicx}                
  \DeclareGraphicsExtensions{.pdf,.png,.jpg,.jpeg} 
\else
  \usepackage{graphicx}                
  \DeclareGraphicsExtensions{.eps}     
\fi%

\usepackage{microtype}                 
\PassOptionsToPackage{warn}{textcomp}  
\usepackage{textcomp}                  
\usepackage{mathptmx}                  
\usepackage{times}                     
\usepackage{cite}                      
\usepackage{tabu}                      
\usepackage{booktabs}                  
\usepackage[usenames, dvipsnames]{color}
\usepackage{enumerate}   
\usepackage[toc,page]{appendix}
\usepackage{hyperref}
\usepackage{kotex}
\usepackage{float}

\onlineid{0}

\vgtccategory{Research}

\title{Thumbnails for Data Stories: A Survey of Current Practices}

\author{Hwiyeon Kim\thanks{e-mail: gnldus28@unist.ac.kr}\\ %
        \scriptsize UNIST %
\and Juyoung Oh\thanks{e-mail: juyoung0@unist.ac.kr}\\ %
     \scriptsize UNIST %
\and Yunha Han\thanks{e-mail: diana438@unist.ac.kr}\\ %
    \scriptsize UNIST %
\and Sungahn Ko\thanks{e-mail: sako@unist.ac.kr (Corresponding author)}\\ %
     \scriptsize UNIST %
\and Matthew Brehmer\thanks{e-mail: mb@mattbrehmer.ca}\\ %
     \scriptsize Independent Researcher %
\and Bum Chul Kwon\thanks{e-mail: bumchul.kwon@us.ibm.com}\\ %
     \scriptsize IBM Research %
     \parbox{1.4in}
     {\scriptsize \centering}
     }

\abstract{When people browse online news, small thumbnail images accompanying links to articles attract their attention and help them to decide which articles to read.
As an increasing proportion of online news can be construed as data journalism, we have witnessed a corresponding increase in the incorporation of visualization in article thumbnails. 
However, there is little research to support alternative design choices for visualization thumbnails, which include resizing, cropping, simplifying, and embellishing charts appearing within the body of the associated article. 
We therefore sought to better understand these design choices and determine what makes a visualization thumbnail inviting and interpretable. 
This paper presents our findings from a survey of visualization thumbnails collected online and from conversations with data journalists and news graphics designers. 
Our study reveals that there exists an uncharted design space, one that is in need of further empirical study.
Our work can thus be seen as a first step toward providing structured guidance on how to design thumbnails for data stories.
} 

\CCScatlist{
  \CCScatTwelve{Human-centered computing}{Visu\-al\-iza\-tion}{}{}
}

\begin{document}




\firstsection{Introduction}
\maketitle

We are witnessing a rise in data-driven storytelling and journalism~\cite{riche2018data}.
News organizations are increasingly publishing articles across a variety of topic areas that are grounded in the analysis of data.
Online news sites such as \textit{The New York Times' Upshot}, \textit{The Pudding}, and \textit{FiveThirtyEight} are specifically devoted to publishing data journalism articles. Such articles are often supported by visualization, presenting annotated statistical graphics, maps, and custom charts to readers.

In this paper, we ask how data-driven articles attract readers. 
Specifically, we consider the role of thumbnails: small and often static images that accompany article titles and bylines, inviting readers to visit the linked article.
Readers may encounter thumbnails while browsing news organizations' home pages, while using a mobile news reader application, or while browsing social media. 
Often these thumbnails incorporate images used in the body of the article in an attempt to convey its gist.
However, when the focus of the article is data and its visual content is predominantly visualization-based, we lack an understanding of how to design an effective visualization thumbnail, one that is inviting, interpretable, and not misleading. 

There is little research to support alternative visualization thumbnail design practices, such as resizing, cropping, simplifying, and embellishing charts appearing within the body of the associated article.
We therefore conducted a survey of visualization thumbnails collected from online news media.
We then had a series of conversations with experts such as data journalists and news graphics designers about their perspectives on designing thumbnails for data stories.
As a result of our survey and conversations with experts, we identify an uncharted design space for thumbnail design, one having little empirical research on which to make effective design choices.
We conclude with a set of future research directions having a common goal of identifying best practices for designing inviting and attractive thumbnails for data stories.

The main contribution of this paper are as follows:
\begin{itemize}
    \item Extracting the key characteristics of thumbnails by surveying a collection of in-the-wild thumbnails,
    \item Deriving thumbnail design strategies and goals based on the survey and conversations with industry practitioners, and
    \item Reporting the lack of consensus on design strategies and goals for visualization thumbnail designs.
\end{itemize}

\section{Related work}
We situate our work in relation to thumbnail design and to the use of visualization for communication, particularly in data journalism. 

\subsection{Designing Thumbnails}

Prior research shows that the presence of thumbnails in search results help people locate articles of interest online~\cite{Aula10,Dziadosz02,Woodruff01}, particularly when paired with informative titles, text snippets, and URLs.
Thumbnails are a particularly important signal of relevance when some links to articles have them and others do not. 
Aula et al.~\cite{Aula10} further showed that presenting thumbnails without accompanying text leads to worse search performance than simply presenting text without a thumbnail.
Thumbnails also trigger behavior that would otherwise not occur in their absence. 
For instance, Topkara et al.~\cite{Topkara12} observed that when an email contains a link to a video file, more people click on the link when it is accompanied by a thumbnail.

Beyond web search and email reading, thumbnails also appear in the context of navigating other forms of media, from file systems~\cite{Robertson98} to documents~\cite{Cockburn06} and videos~\cite{Matejka13}.
Prior work in the visualization and visual analytics community has also incorporated thumbnails into the sensemaking process as a means of leveraging the unique advantages of spatial memory~\cite{Nguyen16, Yoghourdjian18, Heer08}.

Altogether, prior research shows that thumbnails play in helping people \textit{locate} and \textit{rediscover} content during online search, media navigation, and sensemaking.
In contrast, we examine the role that thumbnails play when people \textit{browse} online news articles, and specifically articles falling under the umbrella category of data journalism.

When browsing unfamiliar content, such as in the case of online news reading, thumbnails must compete for the reader's attention with one another and with other content. 
We therefore turn to other prior research examining specific aspects of thumbnail design. 
Several factors appear to impact how thumbnails draw attention and their ultimate utility, such as thumbnail size and the inclusion of text within the thumbnail.
Kaasten et al.~\cite{Kaasten02} studied thumbnail size, concluding that the optimal thumbnail size depends on the task they are intended to support.
For instance, they posit that a thumbnail should be larger than 96x96 pixels in order to trigger recognition among those revisiting a page containing multiple thumbnails.

Other prior research has considered how to automatically generate thumbnails using photos and images appearing in the article, such as by cropping, resizing, or selecting the most salient excerpts from them~\cite{Li08,Suh03}.
More recently, Song et al.~\cite{Song16} introduced an algorithm to select highly salient and evocative thumbnails for videos.
As this research was conducted with generic images, we see our work as a step toward the automatic generation of visualization thumbnails.
This first requires a better understanding of current practices in thumbnail design.

\subsection{Visualization in Data Journalism}

Visualization is increasingly prevalent in news media~\cite{stolper2018}.
In this context, the communicative intent of visualization often leads to different design choices than those used in the context of data analysis~\cite{kosara2016presentation}.
As a result, we encounter substantial use of graphical and text-based annotation~\cite{Ren17}.
We also see the embellishment of charts~\cite{Bateman10,Skau15} with human-recognizable objects~\cite{haroz2015isotype} and the inclusion of graphics not related to the data~\cite{Byrne16}.
These noticeable aesthetic design choices~\cite{Moere12} and additional layers of annotative and embellishment may contribute to positive first impressions with an information graphic~\cite{Harrison15}, and there is evidence to suggest that they increase reader comprehension~\cite{Kong12} and memorability~\cite{Borgo12,Borkin13,Borkin16}.
Hullman and Adar~\cite{Hullman11} argue that while some embellishments make information graphics more difficult to interpret, their judicious application may help readers comprehend and recall content in some cases. 

Designing for first impressions~\cite{Harrison15} and a limited attention span is paramount in the world of online news, where publishers are vying for readers to visit their pages~\cite{Amini2018DDS}.
An unfortunate reality of this news ecosystem is that key messages are often distorted with misleading titles despite accurate visualization content~\cite{Kong18}.
In other cases, deceptive visualization design practices intentionally mislead the reader~\cite{Pandey15}. 
Visualization thumbnails provide readers with a first impression of an article, and when the reader decides not to read the article, the thumbnail will be all they see and potentially remember.
Therefore, it is of utmost importance to design effective thumbnails that are aesthetically pleasing and inviting while being easily interpretable, without being misleading. 
Hence we decided to investigate current practices in visualization thumbnail design.

\begin{figure}[t]
    \begin{center}
    \includegraphics[width=0.8\columnwidth]{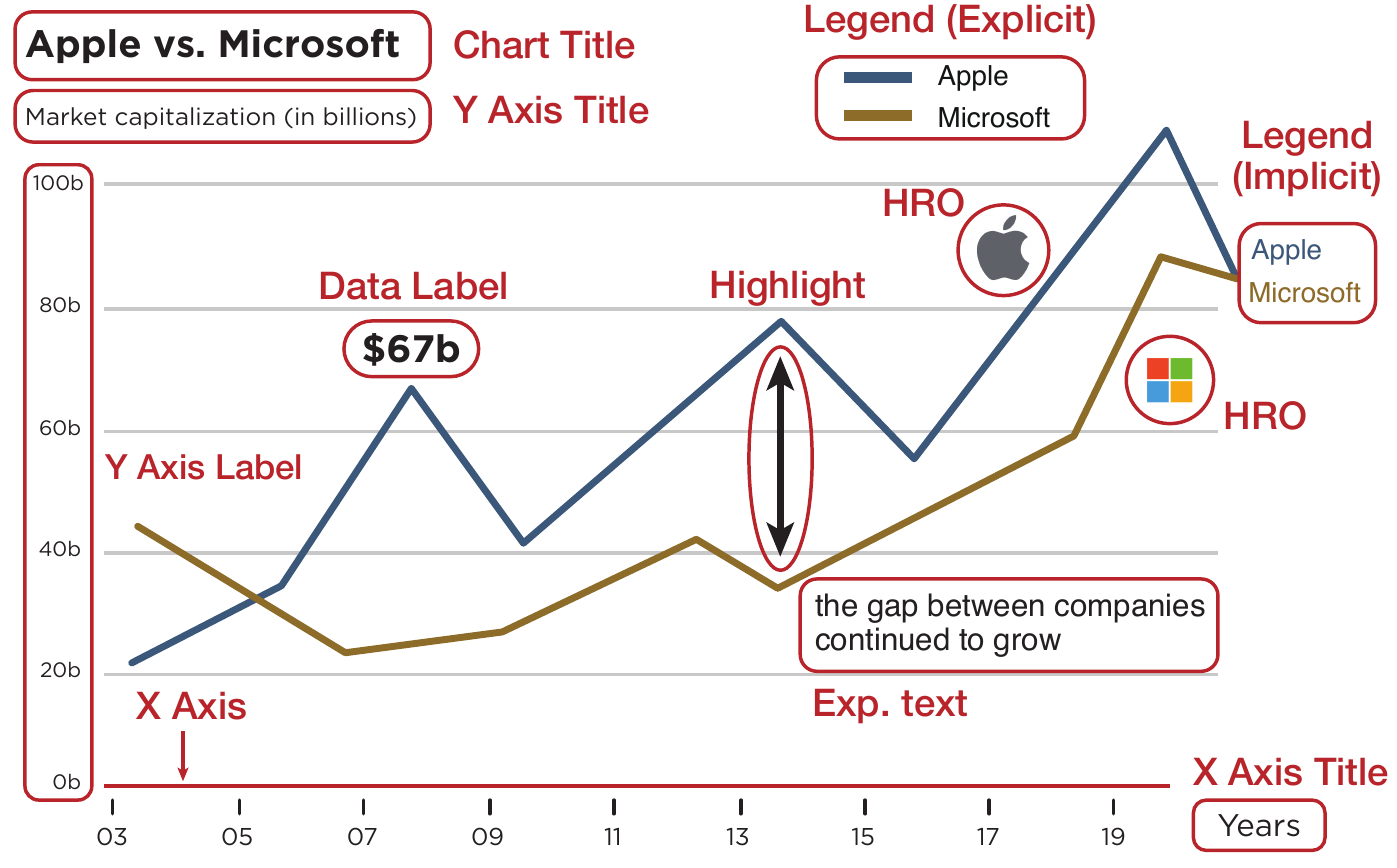}
  \caption{A line chart annotated (in red) according to our classification of chart elements, featuring examples of \textsl{added} components including HROs and highlights, as well as explicit and implicit legends.}
  \label{fig_add_comp}
  \end{center}
\end{figure}

\section{Visualization Thumbnail Design in Practice}
\label{sec_survey}
To our knowledge, prior research has yet to examine the design of thumbnails in data journalism and particularly those which we refer to as \textit{visualization thumbnails}.
To better understand current practices in visualization thumbnail design, we collected examples from data journalism outlets and spoke to news graphics designers.


\subsection{A Survey of Visualization Thumbnails}
We began by collecting news articles published between November 1, 2018 and December 31, 2018 from online news organizations reputed for their data journalism, the \textit{Pew Research Center} (Pew.), \textit{The Economist} (Eco.), \textit{The New York Times, including The Upshot and DealBook} (NYT), \textit{FiveThirtyEight} (538), \textit{The Wall Street Journal (Graphics category)} (WSJ), \textit{First Tuesday Journal} (1st), and \textit{Bloomberg News } (BBG). 
We concentrated on articles relating to politics and economics in graphics categories, as these topics tend to use visualization to a greater degree than others covered by these news outlets. 
This initial corpus contained \textbf{139} articles that included visualization within the article body.
Among these, 48 articles did not feature visualizations in the article's thumbnail, instead opting for photographs and other visual imagery aside from visualization.
Of the remaining \textbf{91}, we decided to focus on basic charts like bar, line, scatter plot, frequently used in news media.
Thus we excluded 24 articles, whose charts are sparingly/infrequently used. 


Among the remaining \textbf{67} articles, 39 included a visualization used in the body of the article as its thumbnail, reducing its size or cropping it. 
Thumbnails for the remaining 28 articles modified a visualization used in the body of the article in some way, such as by omitting axes.
Examining these 67 thumbnails further, we codified 1) which visualization components were modified (e.g., an axis); and 2) how components were modified (e.g., omitted).
The first three authors of this paper independently codified these components and later merged their codes in an iterative discussion, arriving at 96\% agreement (Fleiss' Kappa $=$ 0.75).

To label the components, we considered using existing classifications, namely Borkin et al.'s classification of visualization components~\cite{Borkin16}, Byrne et al.'s~\cite{Byrne16} distinction between graphical and figurative components ~\cite{Byrne16}, and Ren et al.'s~\cite{Ren17} classification of annotation.
We realized that these classifications were insufficient in isolation in terms of capturing all aspects of visualization thumbnail design. 
For example, Borkin et al.'s classification defines `text' as `any text in the image,' so two axis titles, annotations, and captions fall under the same category.
Meanwhile, Byrne et al.'s classification provided broad categories for coding figurative components.
Finally, Ren et al.'s classification could not be used to describe chart components beyond annotation. 
While these classifications informed our analysis, we struggled to use them as a means to codify the designers' intentions or goals. 
We therefore combined and extended the aforementioned classifications, resulting in a new classification having categories that explicitly acknowledge each visualization element's role.





We identify 14 \textit{basic} and 4 \textit{added} component types. 
The basic types refer to chart components that the reader needs to understand the chart (e.g., two axes in line chart).
\autoref{fig_add_comp} shows examples of the basic component types.
They include: x- and y-axes with labels, tick marks, data labels, chart titles, and legends. 
By data labels, we mean any text that directly reflects a data value (e.g., `\$67b' in \autoref{fig_add_comp}).
We further distinguish two types of legends: \textit{explicit legends}: those drawn in a dedicated area; and \textit{implicit legends}: those drawn directly within the visualizations (e.g., Apple \& Microsoft in \autoref{fig_add_comp}). 
Added components include explanation text (or exp. text), highlights (e.g., the vertical arrow in \autoref{fig_add_comp}, the red bar in this thumbnail~
[{\url{https://econ.st/2ZhL5os}}]
, Human Recognizable Objects (HROs)~\cite{Borkin16}), and Graphics Not Relevant to Data (GNRD) in order to capture all forms of graphical embellishment.
HROs are pictorial components used in legends (e.g., the Apple and Microsoft logos in \autoref{fig_add_comp} and a small human object in this thumbnail--
\url{https://bit.ly/2YenA2V}
or to encode data points~\cite{haroz2015isotype}
GNRDs are images or illustrations that reflect the article's context but are not directly related the data, such as the blue image (bottom-right) in this thumbnail
\url{https://bit.ly/2YenA2V}.

\begin{table}[!ht]
    \begin{center}
    \caption{Our classification of 67 visualization thumbnails. Filled values reflect modifications from a visualization found in the article. 
    An interactive version of this table can be accessed at \url{http://hci.unist.ac.kr/vtn_table/html/index.html}}
    \includegraphics[width=0.85\columnwidth]{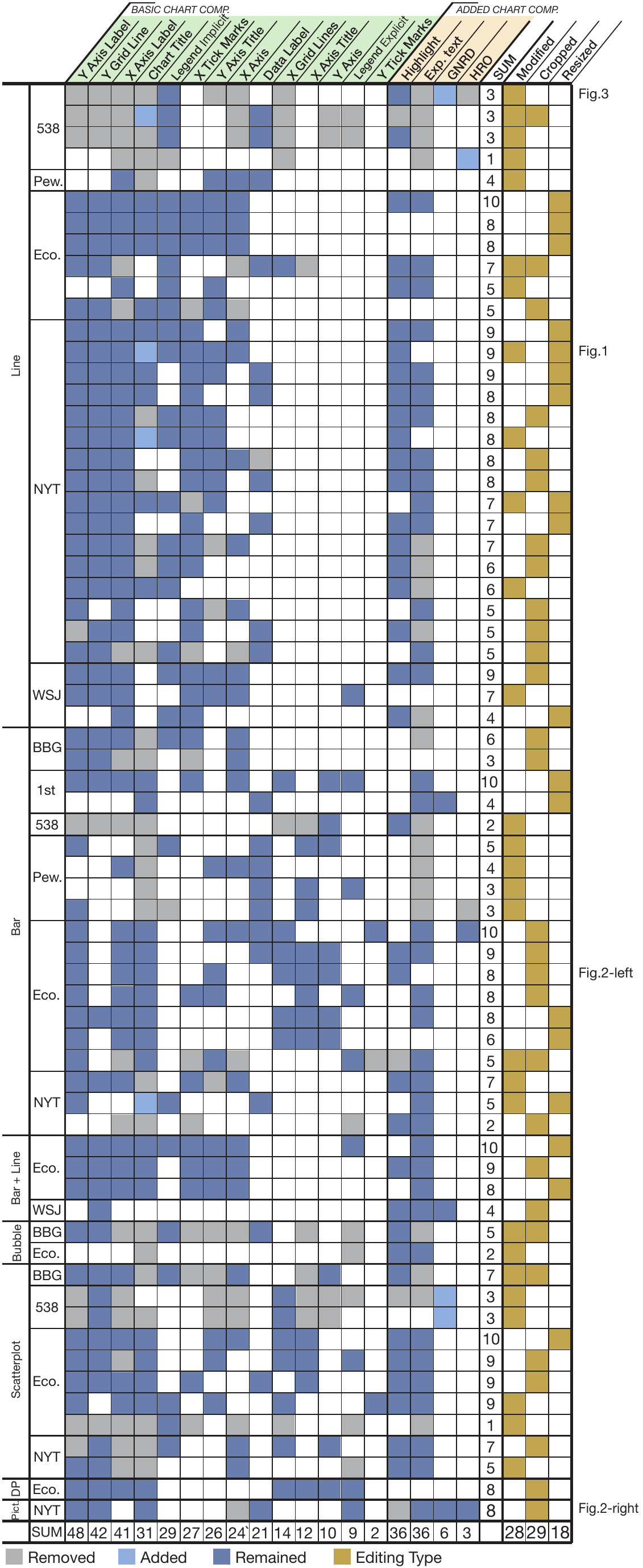}
  \label{tbl_in_the_wild_survey}
  \end{center}
\end{table}

\begin{table}[!ht]
    \begin{center}
    \caption{
    Our classification for additional 24 visualization thumbnails that are not axes-based (e.g., maps) or are less frequently used.} 
    \includegraphics[width=1\columnwidth]{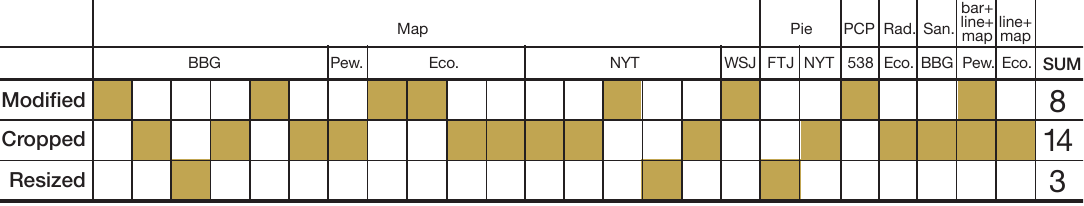}
  \label{tb1_non_trad_charts}
  \end{center}
\end{table}


\autoref{tbl_in_the_wild_survey}\footnote{Online version: \url{http://bitly.kr/AGRYQA}} presents the result of coding 67 visualization thumbnails\footnote{Thumbnail collection: \url{http://bitly.kr/bibRs4}} using our \textit{basic} and \textit{added} component classification, along with our \textit{modified} / \textit{resized} / \textit{cropped} distinction; we also indicate chart types and sources along the table's vertical axis. 
`DP' and `Pict.' in the first column (i.e., chart type) correspond to dot plots and pictograms chart, respectively. 
Links to articles associated with the thumbnails profiled in \autoref{tbl_in_the_wild_survey} are provided in our supplemental material. 

Our codification suggests several trends. 
Thumbnails for line charts (30 out of 67) tend to omit the X-axis title, the Y-axis, and legends.
However, they tend to include added components such as highlights and explanation text.
Other charts, including bar charts and scatterplots tend to include diverse combinations of components that are omitted or added in thumbnails.
We are also able to contrast the strategies of different news organizations.
For instance, \textit{FiveThirtyEight} tends to remove nearly all components from line charts in thumbnails; they also tend to add GNRDs and HROs (e.g.,
\url{https://53eig.ht/330iOEW}).
Meanwhile, traditional print media organizations such as \textit{The New York Times}, \textit{The Wall Street Journal}, and \textit{The Economist} tend to crop or resize charts when producing thumbnails. 
Many media organizations seem to prefer modification (28) and crop (29) strategies in their designs. 
In particular, we see that 538 and Pew are more inclined to modifying, while NYT seems to prefer cropping existing visualizations inside of articles. 
Lastly, we see considerable variability in \autoref{tbl_in_the_wild_survey} which is an indication of a need for greater understanding in terms of how visualization thumbnail design choices affect readers' interpretation as well as readers' likelihood to read the article.

Table~\ref{tb1_non_trad_charts} shows the codification results of the editing strategies (i.e., modified, cropped, or resized) of 24 visualization thumbnails, including maps, pie charts, parallel coordinates, radial column charts (Rad.), sankey diagrams (San.), and composite charts. 
We coded the editing strategies separately for the additional visualizations because those strategies can be applied to any visualizations.
As seen in Table~\ref{tb1_non_trad_charts}, cropping is more popular than the other two strategies for the additional 24 visualizations; however, cropping is not especially popular in the 67 visualizations, which include X-axis and Y-axis, as shown in Table~\ref{tbl_in_the_wild_survey}.
Most maps that are presented as cropped, losing chart titles or legends. 
In analyzing the thumbnails, such a removal seemed effective; it allows more space without causing additional difficulty in recognizing color-coded legends and small text. 
The discussed trends in this section are examples from the tables but are by no means comprehensive.

\subsection{Conversations with News Graphics Designers}
\label{sec_interview}

To understand the current practices in visualization thumbnail design further, we engaged in informal conversations with six visualization practitioners from our extended professional networks who are employed by news media organizations.
These included two journalist-engineers (with 6 and 16 years of experience, respectively), an interactive graphics developer (with 15 years of experience), a senior news article editor (with 12 years of experience), a data scientist (with 6 years of experience), and a computational journalist (with 3 years of experience).
Given the demanding nature of their work and the time zone differences between us, these conversations were asynchronous and occurred via email or within the \#journalism channel of the Data Visualization Society's Slack workspace\footnote{\url{https://datavisualizationsociety.com}}.


We asked these practitioners about two topics: (1) their intentions with respect to designing thumbnails for articles that prominently feature visualization; and (2) the challenges of incorporating visualization into article thumbnails. 
These practitioners reported a broad set of goals for thumbnails. 
First, their thumbnails must build and reinforce the organization's brand identify.
This requirement often constrains the choice of colors, font types, and HROs used in thumbnails.
Second, their thumbnails must be aesthetically appealing in order to draw readers' attention, particularly when the thumbnail appears in a distracting social media feed.
Third, their thumbnails must also reflect any unique artwork or visual content commissioned for the article (if applicable), which might include illustrations, collages, or animations.

In regards to how they design thumbnails for data stories, these practitioners indicated that there are ``\textit{no hard and fast rules of thumb}'' that can be applied to all cases.
Aside from using a limited color palette to reinforce brand recognition, two of these practitioners admitted to avoiding the use of visualization in thumbnails.
Instead, they opted to incorporate photographic imagery whenever possible, and that photos of people appear to drive more traffic to articles.
Maps also appear to successfully drive readers to articles.  
One practitioner reported that visualizations are often regarded by readers as ``\textit{cold, intimidating, or inaccessible},'' and that while a visualization in isolation does not communicate much in itself, they can sometimes be used to accentuate a photographic or illustrated thumbnail. 
For instance, a composite thumbnail can incorporate a photo of a person with a semi-transparent visualization overlay.  
This practitioner also avoided incorporating visualization in thumbnails because they did not want to ``\textit{give away too much content}'' before the reader arrives at the article. 
Lastly, we also learned that creating a thumbnail for an article is often not the responsibility of the article author or visualization designer. 
Instead, thumbnail design is designated to designers and social media content producers who are typically not involved in writing the article.

Though our conversations with practitioners were informal and by no means exhaustive, we were encouraged to learn about the lack of consensus in regards to guidelines or patterns for visualization thumbnail design.
We therefore remained curious about what makes visualization thumbnails effective, or how their components contribute to whether readers find them to be inviting and interpretable.
Conversely, we also questioned which thumbnail components could be used to mislead or misinform readers.
Considered alongside our survey results, these conversations reinforced the need for further empirical study.

Based on the survey results and our conversations, in this paper we consider a \textbf{visualization thumbnail} as \textit{a thumbnail that provides an article's overview through visualizations to invite readers to click on the thumbnail for further reading.} 
Two perspectives of this definition exist, and each perspective leads to different visualization thumbnail design goals. 
From the perspective of a professional designer, whose ultimate goal is to draw readers' attention to the thumbnails and increase thumbnail clicks, visualization thumbnail goals are not much different from those of non-visualization thumbnails.
Therefore, conventional design goals could work for visualization thumbnail evaluation, such as determining its attractiveness~\cite{Song16} or visual aesthetics~\cite{Moere2011, Harrison15, Cawthon2007}. 

However, from the perspective of a reader, whose ultimate goal is selecting articles that best match his or her subjective criteria (e.g., preferences, interests, or intentions), whether a thumbnail is attractive or visually pleasing might not be the most important criteria for selecting that thumbnail.
Rather, the most relevant criteria for thumbnail design might involve making a visualization thumbnail that would allow readers to judge accurately and quickly whether the thumbnail’s article meets their criteria.
This perspective leads to further design goals, such as informativeness, relevance, interpretability, and straightforwardness ~\cite{Woodruff01, Aula10, Lam05, Li08, Teevan09}. 
In conclusion, we think that visualization thumbnails can be designed according to the different goals of designers and consumers.  
While some goals of producers and consumers may compete with one another (e.g., visualizing a full story may end up with no clicks vs. providing little information on the story may cause readers to feel misled), we think that there is a trade-off point at which visualization thumbnails not only successfully invite consumers with the article led by the thumbnail but also contribute to increasing clicks on the thumbnails. 
Our work is the first attempt on this topic. Due to the limitations in the existing methods and scopes of the study, the proposed definition should be considered as a working definition and could evolve or be replaced with the results of follow-up experiments and studies.

\section{Lesson Learned and Future Work}
At the outset of this project, we began by asking what makes thumbnails for data stories inviting and interpretable.
We conducted a survey of visualization thumbnails and had a series of conversations with practitioners about the design of thumbnails for data-driven stories. Our study results reveal a design space for thumbnails, a set of thumbnail design components that can be leveraged to attract readers and help them to understand the main point and context of articles associated with thumbnails.
We then 
asked six practitioners from news organizations to reflect on the use of thumbnails in their own organizations.
Ultimately, our studies shed a light on an uncharted design space for visualization thumbnail design and toward automatically generating or recommending an ideal set of visualization components to include in a thumbnail.

As future work, we intend to further examine readers' first impressions of thumbnails, considering both their visual aesthetics and informational aspects.
Example research questions consider the relationship between titles and Graphics Not Related to Data (GNRDs).
We will consider cases where GNRDs may stand alone as well as cases in which they hinder interpretation. 
Animated visualization thumbnails~\cite{Leiva13} is also an exciting potential area of future study. 
As thumbnail sizes are a strong constraint, we plan to investigate the impact of different sizes on the effectiveness of thumbnails.

\acknowledgments{
The authors wish to thank all journalists who provided constructive feedback for our project.
This work was supported by the National Research Foundation of Korea (NRF) grant funded by the Korea government (MSIT) (No. 2017R1C1B1002586).
}

\bibliographystyle{abbrv-doi}

\bibliography{bib/TN_Gen,bib/Chartjunk_Aesthetics,bib/Enhancement_tech,bib/minimalism_Data_ink_ratio,bib/NarrativeVis,bib/TN_App,bib/TNV_FW,bib/Social_Data_Analysis,bib/Elicitation_Study,bib/Vis.bib,bib/Visualization_literacy}
\end{document}